\documentclass[conference]{IEEEtran}

\usepackage{graphicx}
\graphicspath{{pictures/}}
\usepackage{float}
\title{\textbf{ Immutable Digital Recognition via Blockchain}
}
\author{
    \IEEEauthorblockN{Zeng Zhang}
    \IEEEauthorblockA{\\Hainan University \\ Haikou China\\ \textbf{zz1up@hainanu.edu.cn}}
    \and
    \IEEEauthorblockN{Xiaoqi Li}
    \IEEEauthorblockA{\\Hainan University\\Haikou China\\\textbf{csxqli@ieee.org}}
}
\date{\today}

\renewenvironment{abstract}
 {
  \begin{center}
  \bfseries \abstractname\vspace{-.5em}\vspace{0pt}
  \end{center}
  \list{}{\rightmargin\leftmargin}
  \item\relax}
 {\endlist}

 \usepackage{titlesec}

\titleformat{\section}{\normalfont\Large\bfseries}{\thesection}{1em}{}
\titleformat{\subsection}{\normalfont\large\bfseries}{\thesubsection}{1em}{}

\renewcommand{\thesection}{\arabic{section}}
\renewcommand{\thesubsection}{\arabic{section}.\arabic{subsection}}

\titleformat{\section}[block]
  {\normalfont\large\bfseries}{\thesection}{1em}{\MakeUppercase}

\begin{document}

\maketitle

\begin{abstract}
The development of blockchain technology has significantly enhanced the security and transparency of personal information and transaction records. Concurrent with the advancement of blockchain technology and the emergence of the digital currency ecosystem, the internet has evolved from a paradigm dominated by information flow to one driven by value flow. Consequently, the concept of token has gained widespread dissemination, and the electronic token under investigation in this thesis is a development of this concept. The application of electronic tokens has become pervasive with the development of the internet, but the functionality of these tokens is often limited, and issues related to trust remain significant challenges. This study proposes innovative solutions to address the deficiencies in traditional electronic token systems, including the issuance of tokens, the heterogeneity of issuance standards, and the lack of democracy. The solutions are based on distributed storage, anti-tampering mechanisms, consensus protocols, and transparent, traceable storage. Additionally, the study employs NFTs, smart contracts, and RSA blind signatures, among other key technologies, to construct a system based on blockchain technology. The study's findings have led to the identification of a centralised institution that ensures authority. The architecture of the electronic certification system is characterized by a combination of centralized and decentralized components. In order to achieve a harmonious balance between authority and innovation, this paper proposes a rigorous certification process. The process integrates the \textbf{decentralised management} and \textbf{centralised operation} models, aligning them with the national policy directives. The developed solution enables the full utilisation of blockchain technology's advantages while also fostering community participation. Consequently, it establishes a secure, legal, reliable, and dynamic electronic certification system.
\end{abstract}

\section{INTRODUCTION}

The advent of the information era and the rapid advancements in blockchain technology have rendered digital honorary medals increasingly popular among the market and the general public. These digital certificates and awards have been widely adopted in various domains
\cite{cobuccio2025medal},
such as social interaction, identity verification and public relations, due to their versatility in showcasing personal qualifications and status. However, the traditional digital honorary medal system is typically administered and managed by centralised institutions such as government entities or large corporations. This has resulted in several issues, including the lack of uniformity in issuance standards, the complexity of management processes, the vulnerability of personal information disclosure, limited user participation and the inequitable selection process. To address these issues, we have employed blockchain technology in the digital medal system, thereby establishing a more efficient, transparent and equitable honor system.

Distinctive characteristics of blockchain technology include its decentralised nature, the integrity of its data, and the ability to trace the history of transactions
\cite{liu2022blockchain}\cite{deshmukh2022blockchain}\cite{angelis2019blockchain}.
These characteristics align with the technical standards of electronic certification systems . The establishment of electronic certification systems on blockchain technology ensures the immutability of certified information. It also provides a transparent and open process for the issuance and verification of certifications.To facilitate the supervision of the general user base, the system has been developed to include both a smart contract-based automatic issuance function and a blockchain-based electronic voting function
\cite{mohanta2018overview}.\cite{zou2019smart}
This allows the electronic token to be distributed through a public vote
\cite{wang2018large},
thereby enhancing the credibility and reach of the distribution platform, increasing the effectiveness of the electronic token's promotional efforts, and achieving its intended purpose of propagating a specific value.

By the prevailing policy directives of the nation, the establishment of a fully decentralised system is not legally recognised. Consequently, the present project will utilise the technical aspects of blockchain and electronic seals to achieve a balance between \textbf{centralised control} and \textbf{decentralised dynamism}. The objective is to construct a national electronic seal architecture that is governed by the nation, with contributions from multiple platforms
\cite{hou2017application}.
\section{BACKGROUND}
\subsection{Traditional Electronic Medal}

\subsubsection{\textit{ Current Situation}}
Currently, electronic medal system is developing rapidly, but there are still some issues with the medal awarding scheme and certification system:

(1) The awarding mechanism is too broad

Currently, electronic medals are mostly awarded in a ‘wholesale’ manner, with a group of recipients being pre-selected and then awarded en masse. However, this awarding logic leads to two problems: on the one hand, the awarding criteria of this system cannot recognise individual contributions, making it difficult to achieve personalised recognition of the recipients. On the other hand, the excessive issuance of medals dilutes their symbolic value, resulting in an inability to create effective personalised incentives while also weakening their role as a symbol of honour.

(2) Lack of uniform technical standards 

Currently, there is no unified standard for digital medal management in China. Searches reveal that different departments implement medal awarding in different ways (some distribute medals via WeChat mini-programs, some use email push notifications, and others develop independent applications), which also makes it difficult to share medal data. This systemic flaw has resulted in two issues: first, electronic medals lack authoritative certification, forcing recipients to rely on physical certificates to prove the authenticity of their medals; second, the absence of a central database shared among departments prevents mutual recognition of medal information, thereby severely undermining the efficiency that digital management should inherently provide. 

(3) Insufficient public participation

The current electronic medal awarding mechanism is significantly department-led, with standards set and recipients selected internally by the department, lacking public input and social oversight channels. This behaviour has severely undermined the credibility of electronic medals.

\subsubsection{\textit{ Issuance process}}

The awarding of traditional medals is typically handled by the government or key corporate institutions, and the process generally involves the following steps: First, individuals or organisations submit the relevant application materials to the appropriate institution. Second, the institution manually reviews the applicant's qualifications. Third, after approval, the physical medal is produced and created. Fourth, following the distribution of the medal, the awarding process information is recorded in the institution's internal database. Fifth, if necessary, verification can be conducted by contacting the awarding institution for assistance.

\begin{figure}[htbp]
  \centering
  \includegraphics[width=\linewidth]{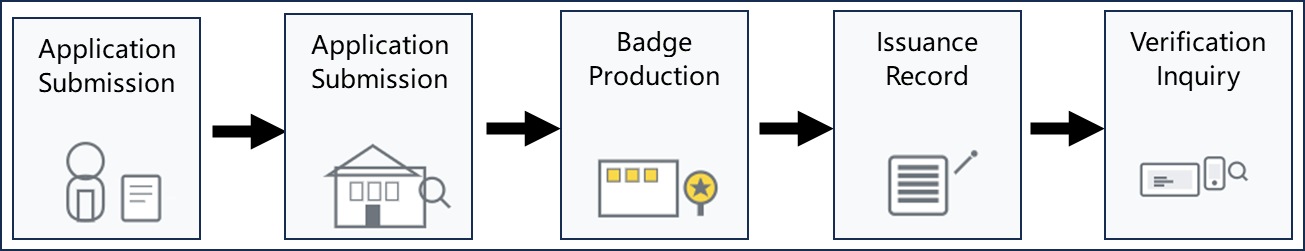}
  \caption{Traditional medal awarding process flow chart}
  \label{fig:medal-process}
\end{figure}
\subsection{Blockchain Technology }

Blockchain is a combination of multiple existing technologies. Data blocks are linked in chronological order and cryptographic techniques are used to ensure that they cannot be tampered with and that transaction data is traceable
\cite{yaga2019blockchain}
. Its decentralised distributed ledger technology is achieved through the joint participation of various nodes in storing transaction records. Below, we will introduce the four most important technical features of blockchain
.

(1) Distributed Storage

In traditional systems, data is typically stored on a specific central server. The advantage of this storage method is that it is easy to manage, but the disadvantage is that it poses certain security risks. If the central database is attacked, it is easy for data to be tampered with, and it may be difficult to detect even after the data has been tampered with. Blockchain, however, employs a decentralised distributed storage approach—it distributes data across various nodes within the network. Each blockchain user can act as a node, and each node participates in data storage and maintains a complete copy of the data. The advantage of this approach is that even if some nodes fail or are attacked, the majority of unaffected nodes on the blockchain network can continue to operate normally. This enables the formation of a truly decentralised distributed storage architecture
\cite{wang2022distributed}
.

\begin{figure}[htbp]
    \centering
    \includegraphics[width=0.5\linewidth]{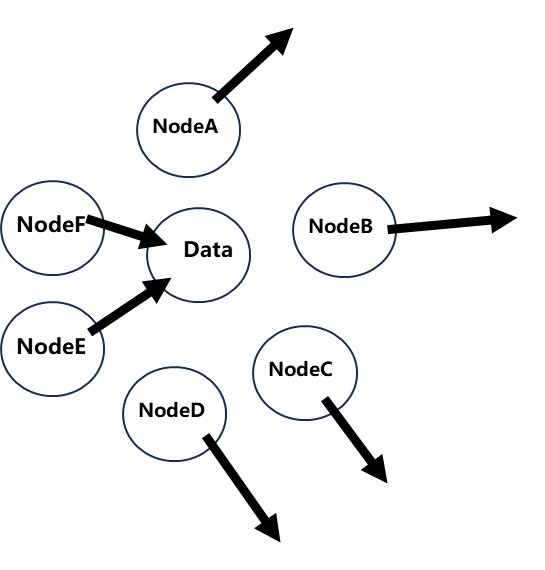}
    \caption{Decentralized distributed storage }
    \label{fig:placeholder}
\end{figure}
(2) Data Anti-Falsification and Anti-Tampering

The blockchain's ability to achieve its highly tamper-resistant nature relies on two core technologies: Merkle trees and cryptographic hashing. Transaction data within each block in the blockchain is linked via Merkle trees. As shown in the figure below, a Merkle tree resembles an inverted tree. Each leaf node on the Merkle tree represents a transaction. Each transaction is hashed to produce a hash value, which is then combined in pairs and hashed sequentially, ultimately resulting in a root hash value. Any minor change to the transaction data will result in a significant change to the root hash value. Therefore, we can use these two technologies to achieve data anti-counterfeiting and anti-tampering in blockchain
\cite{fei2024research}. 

(3) Consensus Mechanism

In the decentralised organisational structure of blockchain, we have adopted the consensus mechanism technology to ensure that all nodes reach a consensus on data consistency. Taking Bitcoin as an example, the consensus mechanism it uses is Proof of Work (PoW). This mechanism requires all nodes on the network to solve a complex mathematical problem to prove their work effort. Only nodes that successfully solve this problem can obtain and possess the right to record transactions
\cite{lashkari2021comprehensive}
.

In Bitcoin, the term mining refers to this process. Mining nodes continuously attempt different random numbers until they find a solution that satisfies the specific conditions for the block hash, thereby obtaining the right to record transactions . This design ensures that only nodes capable of providing sufficient computational power have the opportunity to obtain the right to record transactions, thereby further preventing malicious nodes from tampering with blockchain data
\cite{judmayer2022blocks}.
\begin{figure}[H]
    \centering
    \includegraphics[width=1\linewidth]{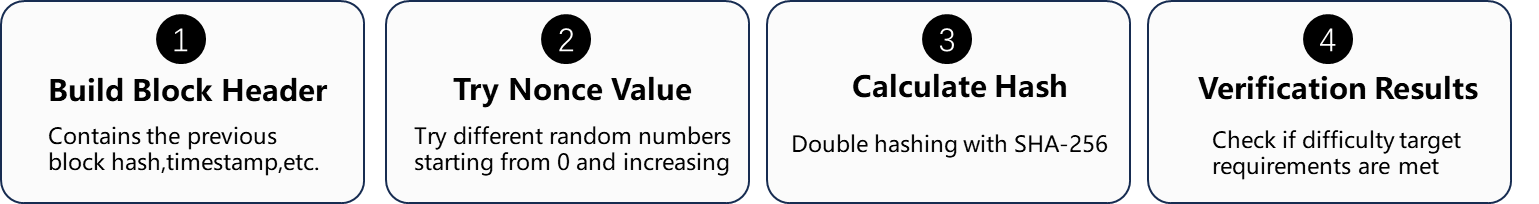}
    \caption{Mining calculation process }
    \label{fig:placeholder}
\end{figure}
(4)Traceable

One of the key characteristics of blockchain technology is its public transparency and traceability. In a public blockchain environment, all transaction records are open to the public, and any individual has the right to query and verify such transactions. Each transaction is accompanied by a unique hash value and timestamp. This information is processed through multi-layer hashing in a Merkle tree, forming a continuous historical chain. This mechanism ensures that every transaction on the blockchain can be precisely tracked and traced
\cite{hastig2020blockchain}.

These characteristics ensure that every transaction within the system is traceable. Users have the right to query the complete transaction history of any transaction and use the hash value to verify the authenticity and legality of the transaction.
\subsection{ NFT}

NFT, or non-fungible token, is a unique digital certification method characterised by its irreplicability and absolute uniqueness. Each NFT is unique, similar to issuing an irreplicable ‘ID card’ for items or events on the Internet
\cite{wang2021non}
. For example, suppose person A creates a painting and turns it into an NFT to send to person B. Even if person B can replicate this NFT,  the copied version differs from the original. When Person C receives the copied version, they can immediately identify it as not being the original. The NFT grants this photo the ability for Person C to verify its authenticity. The NFT's ‘identity card’ information is recorded on the blockchain, ensuring that the information cannot be forged or altered. It is similar to collecting stamps or commemorative coins, each of which has its own unique serial number and identifiable authenticity characteristics. Today, many people use NFTs to represent digital artworks, game items, or various certificates and medals. For example, if you pass an exam or complete a task, the system will award you an NFT medal, which belongs solely to you and cannot be taken away or replicated by others. You can also take it to other platforms to display or use it. NFT technology addresses the issues of traditional digital certificates being easily lost or difficult to verify. The information contained in an NFT includes its content, owner, creation time, etc., and all this information is securely stored on the blockchain. 

It is precisely because of these characteristics that NFTs have great application value in areas such as digital collections, gaming and entertainment, and identity verification. They also provide a new solution for the issuance and storage of blockchain-based electronic badges, which is the focus of this study
\cite{guidi2023nft}.

 \begin{figure}[htbp]
     \centering
     \includegraphics[width=1\linewidth]{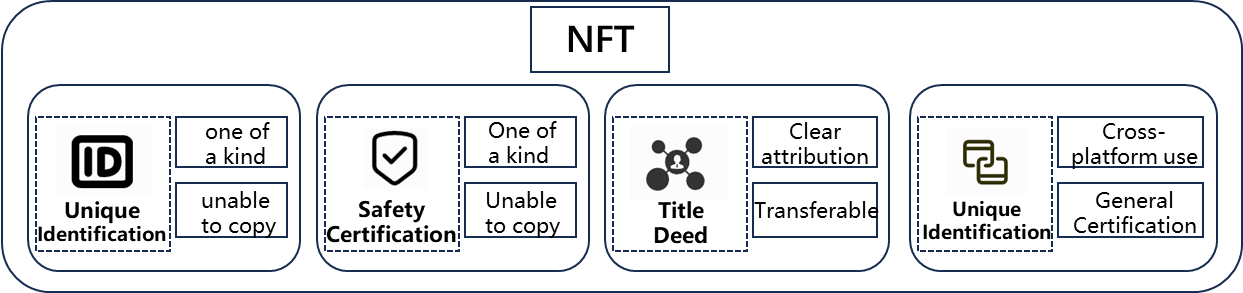}
     \caption{NFT introduction visualisation }
     \label{fig:placeholder}
 \end{figure}

\section{ IMPLEMENTAION}

\subsection{ System Overview }

The electronic badge system is an intelligent digital certification system. Users can obtain virtual badges certified by the blockchain or a centralised database by completing specific goals or going through the corresponding badge issuance process on the blockchain. Types of electronic badges include blockchain NFT badges, centralised database badges, cross-platform achievement systems, and hybrid certification mechanisms (off-chain behaviour + on-chain rights confirmation). Their certification and display forms can cover various application platforms on the Internet.

Based on the interaction scenarios and certification levels of badges, the electronic badge system can be divided into two categories: The first category is platform-level badges: these are automatically obtained by completing predefined tasks within a specific platform (such as an education system, game, or community) and can only be displayed and used within that platform. The second category is certification-level badges: after verification by a centralised authority, achievements across multiple platforms are consolidated into an authoritative database, generating a blockchain-based digital certificate valid across all platforms. For example, if an individual meets the voting criteria, initiates a vote, and satisfies the badge issuance requirements, the corresponding badge is awarded to that individual, and an electronic badge is generated that can be displayed on all platforms, with the achievement also recorded in the government database.

\subsection{Security Implementation}

Generally speaking, to ensure the security of an electronic badge system, the following core requirements must be met. Only by meeting these requirements can the security of the badge system be fundamentally guaranteed. The security requirements of the system can be summarised as the following main features:

• Authenticity: Each badge must be issued by a legitimate authorised institution.

• Uniqueness: Each medal must be awarded to a unique recipient. 

• Immutability: Once award records are uploaded to the database, they cannot be modified. 

• Verifiability: The authenticity of medals must be publicly verifiable. 

\subsection{Combining Blockchain Technology}

Blockchain has four main characteristics: 

(1) Decentralised storage: Medal records are distributed across the blockchain network. 

(2) Data tamper-proofing and anti-forgery: Hash chains are used to ensure data cannot be tampered with. 

(3) Data transparency and traceability: All medal issuance records are publicly accessible and traceable. 

(4) Smart contracts: Enable automated issuance and management of medals.
The inherent characteristics of blockchain technology perfectly align with the security requirements of a badge system. We have integrated the key features of blockchain with the badge system we are developing to ensure the security of the badge system.

\begin{figure}[H]
    \centering
    \includegraphics[width=1\linewidth]{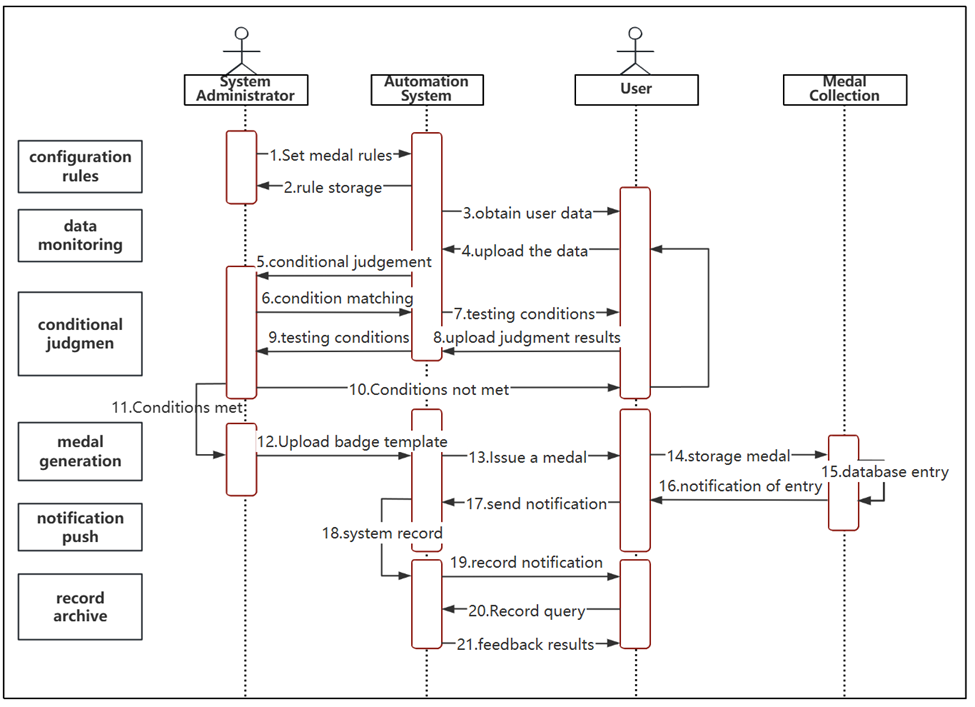}
    \caption{Electronic medal issuance flowchar}
    \label{fig:placeholder}
\end{figure}
\subsection{Combining Smart Contract}

\subsubsection{\textit{ Smart Contract}}

A smart contract is a digital, automatically executed agreement deployed on a blockchain. Essentially, it converts traditional contract terms into automated programs enforced by code. It acts as an autonomous digital arbitrator, using pre-set rules and logic (such as when a user achieves a specific milestone, automatically generates and distributes a badge) to handle corresponding transactions on the chain\cite{mohanta2018overview}.
\begin{figure}[htbp]
    \centering
    \includegraphics[width=0.5\linewidth]{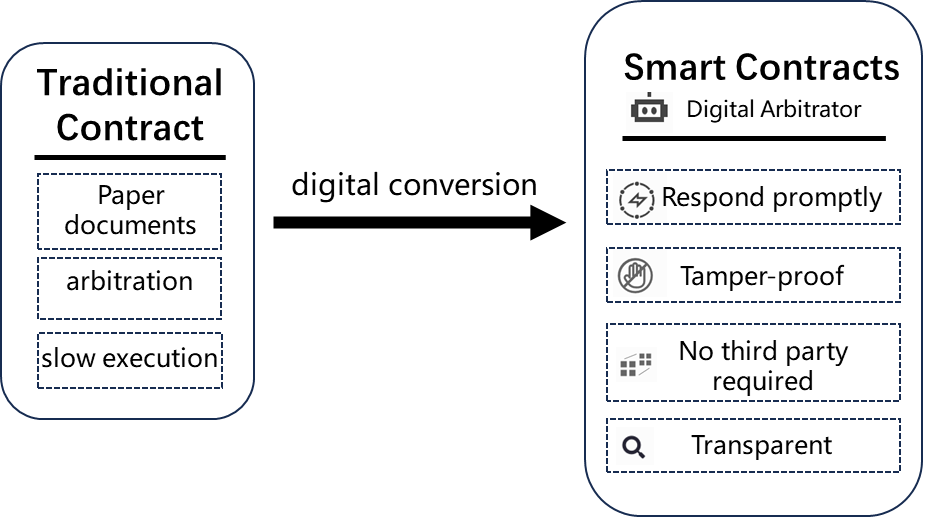}
    \caption{Smart contracts as digital arbitrators }
    \label{fig:placeholder}
\end{figure}

When the predefined conditions are met, the smart contract automatically executes a series of predefined operational processes (such as minting NFT badges, updating ownership records, etc.) . A smart contract is a computer program that can automatically execute predefined agreements, first proposed by Nick Szabo in the 1990s 
\cite{zheng2020overview}. Smart contracts replace the role of transaction contracts without requiring supervision by a trusted third party . In 2017, scholars began to increase their focus on the application of smart contracts in transactions, and related research findings gradually became more abundant . In the study of blockchain-based electronic medal systems, smart contracts can serve as core technology to achieve automated management, handling the generation and issuance of electronic medals, and leveraging blockchain's immutable characteristics to ensure transparent and trustworthy processes. Additionally, smart contracts can retain regulatory interfaces to enable government agencies to exercise ultimate control.

The technical implementation of smart contracts in a badge system may include: (1)Condition- triggered contracts that automatically verify user behaviour and generate and award badges when users meet predefined conditions. (2) Community governance voting contracts that adjust badge award rules based on community user feedback. (3) Cross-chain intera
ction contracts, enabling blockchain badges across multiple platforms to synchronise badge data. (4) Permission control contracts, allowing governments/institutions to retain necessary emergency freeze and revocation rights
\cite{lin2022survey}.
\begin{figure}[htbp]
    \centering
    \includegraphics[width=0.5\textwidth, height=8cm, keepaspectratio]{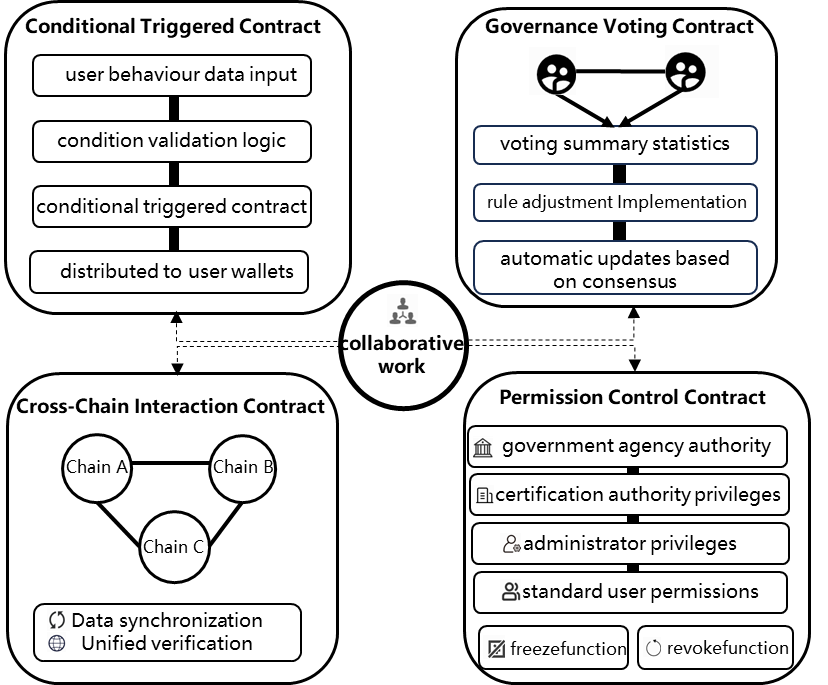}
    \caption{Detailed explanation of the four major technologies of smart contracts }
    \label{fig:placeholder}
\end{figure}

\subsubsection{\textit{\textbf{Realisation}}}

The badge system of the integrated architecture can be divided into two categories based on the functional hierarchy of the contracts:

The first category is execution-level contracts: when users meet predefined conditions (such as completing learning tasks or passing skill exams), badges are automatically created by smart contracts without manual review and awarded to users' platform accounts. The second category is governance-level contracts: These are controlled by centralised institutions (such as government agencies) and are used to review the aforementioned certification-level badges, adjust issuance rules, and handle exceptional appeals, while retaining certain manual intervention permissions.

Using smart contracts, we can automate certain processes. By integrating them with the badge system, we can achieve a complete architecture for a badge system that combines centralised supervision with decentralised operations. This article will use diagrams to describe the overall architecture of this badge system integrated with smart contracts, as well as the overall process for awarding badges within this architecture.

\begin{figure}[htbp]
        \centering
        \includegraphics[width=1\linewidth]{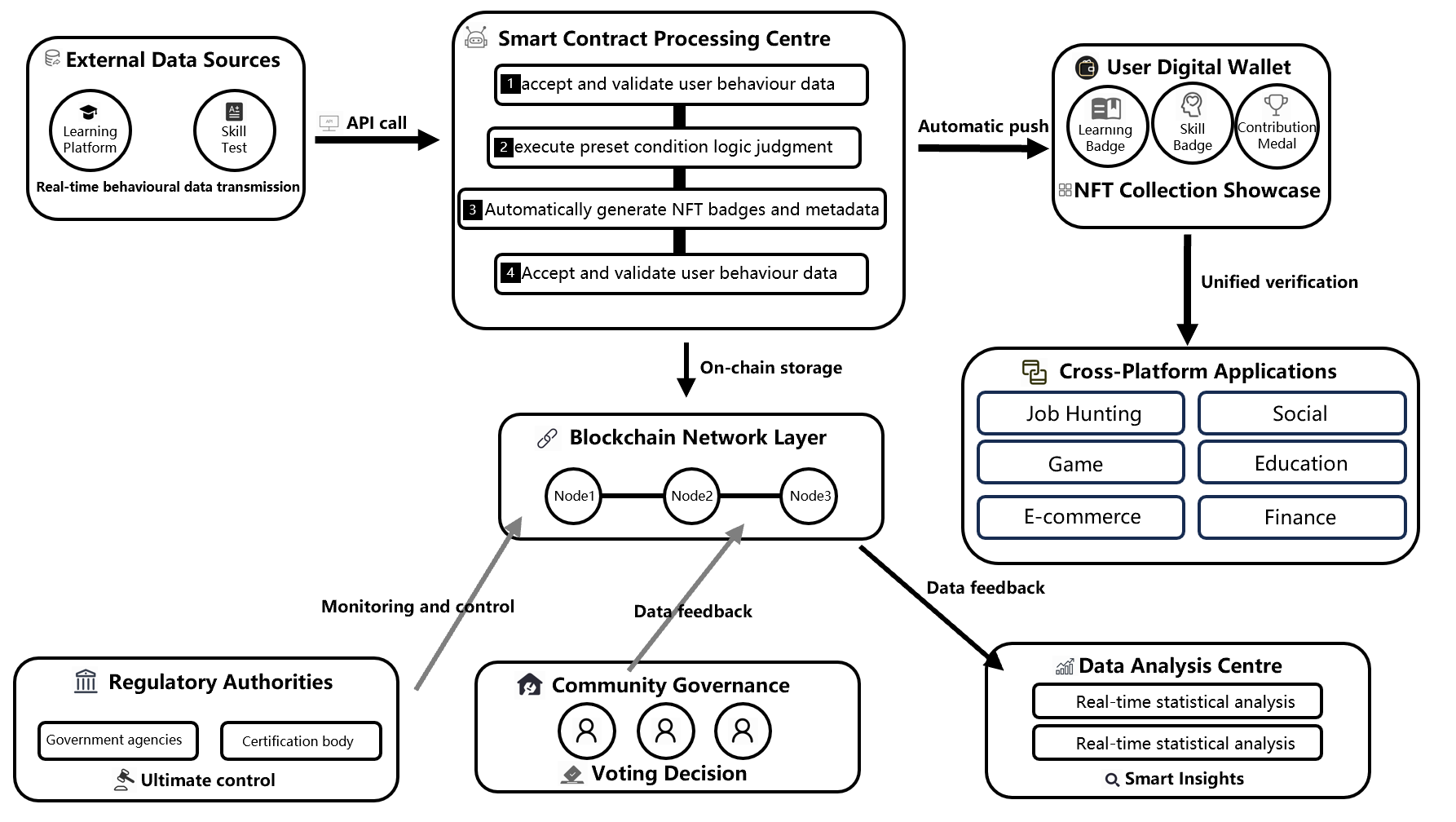}
        \caption{Complete workflow of the smart contract badge system}
        \label{fig:placeholder}
    \end{figure}

\section{CONVERGED ARCHITECTURE SYSTEM}

To meet national policy requirements while leveraging the technical advantages of blockchain technology and stimulating user engagement, we have designed a hybrid blockchain-based electronic medal system that combines ‘centralised institutional authority assurance with decentralised dynamic operations.’ This system not only ensures government oversight of the medal system to enhance its authority and compliance but also stimulates user engagement across multiple platforms to enhance the system's vitality and innovation. Through this system, we can effectively integrate a blockchain-based electronic voting system to ensure the credibility and authority of the medals issued.

\subsection{ Core Architecture}

Centralised Authoritative Database (Foundation):

Entity: Government departments designated by the state (such as the Ministry of Industry and Information Technology, Ministry of Education, Ministry of Culture and Tourism, etc.) or large state-owned enterprises/institutions that have been rigorously selected and possess extremely high credibility and trustworthiness are responsible for management and operation. Core responsibilities include:

Defining the format standards for badges (e.g., name, icon, description, award criteria framework, grading system, etc.) and establishing uniform standards. Serving as the final authoritative repository for all verified and certified badges. Only badges entered into this database possess official recognition and achieve final certification and storage. Generate a unique, blockchain-based identifier (e.g., NFT) for each certified medal to ensure its immutability, traceability, and ownership information, forming a unique identifier. The medal's key data (e.g., unique ID, name, official description, certification status, issuer information, holder address, etc.) is stored on the blockchain to ensure transparency and immutability. The central department will conduct final compliance, security, and value reviews of the medals submitted for application, implement strict review and supervision, and establish overall management strategies, security standards, and audit mechanisms for the medal ecosystem. It will oversee the behaviour of all participating platforms and refine strategy formulation and supervision. 
\begin{figure}[H]
    \centering
    \includegraphics[width=1\linewidth]{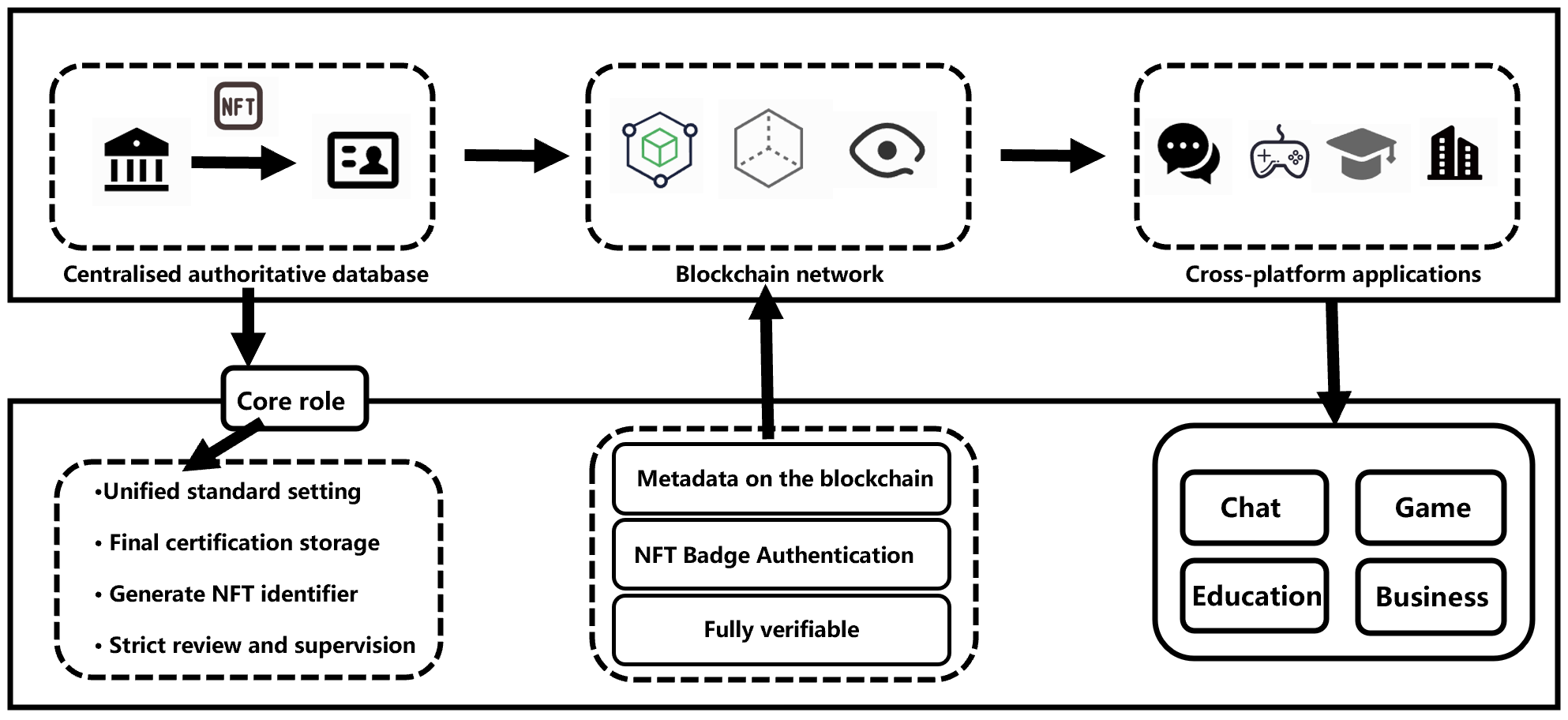}
    \caption{Blockchain-based electronic medal system with integrated architecture}
    \label{fig:placeholder}
\end{figure}

Decentralised Platform Operations (Core): Participating entities: Various qualified enterprises, organisations, and community platforms (such as social media, gaming platforms, educators, industry associations, and even individual developers/creators within an authorised framework). Core responsibilities are as follows:

Platforms design specific badge styles, detailed acquisition conditions (challenges, tasks, achievements, learning objectives, etc.), and incentive mechanisms based on their own business operations, user interests, or specific objectives, while adhering to the metadata framework established by the central database. They then establish badge design and proposal criteria. Two badge issuance formats are adopted: The first involves users automatically triggering the badge acquisition process by completing pre-set tasks, challenges, learning paths, or community contributions on the platform, with badges automatically issued upon successful review. The second introduces a key decentralised element: for certain badges (especially those involving subjective evaluations such as social contributions, promotional effects, or influence), the platform can design a voting mechanism. A professional panel composed of platform users or public voters will vote on user applications, immediately award electronic badges, and upload the data to the blockchain.

\subsection{Core Processes}

If a medal is successfully awarded on a platform and the platform deems it to be valuable, representative, and compliant with the standards specified by the central database, the platform may submit an application for certification of the medal to the central database. The application must include details of the medal, awarding rules, examples of medals already awarded (user addresses, proof of eligibility, etc.), community voting data (if blockchain-based electronic voting technology was used), and other supporting materials. After the application is submitted on the platform, the central database will conduct a rigorous review in accordance with regulations to ensure that the medal's content, name, and acquisition methods comply with national laws and regulations, socialist core values, and central database policies. It will also reassess whether the medal's design is reasonable, whether the acquisition conditions are clear, fair, and valuable, and whether they truly reflect users' contributions and achievements. Further verification of the platform's qualifications will be conducted to determine the authenticity and reliability of the submitted award records, voting data, etc. Finally, the review assesses whether the badge and its mechanism have any security vulnerabilities or potential risks. If the review is approved, the central database will officially record the badge (and its related data) in the authoritative database and generate an official, blockchain-based unique identifier (NFT) for it. User records previously issued by the platform that meet the certification badge standards will also be linked to this official database, and users will automatically obtain the officially certified badge version. If a badge receives official endorsement, it can be used, displayed, and verified across the entire network ecosystem. If the review fails, the central database will provide feedback on the reasons, and the platform may choose to modify the information and resubmit or withdraw the application.
\begin{figure}[H]
    \centering
    \includegraphics[width=1\linewidth]{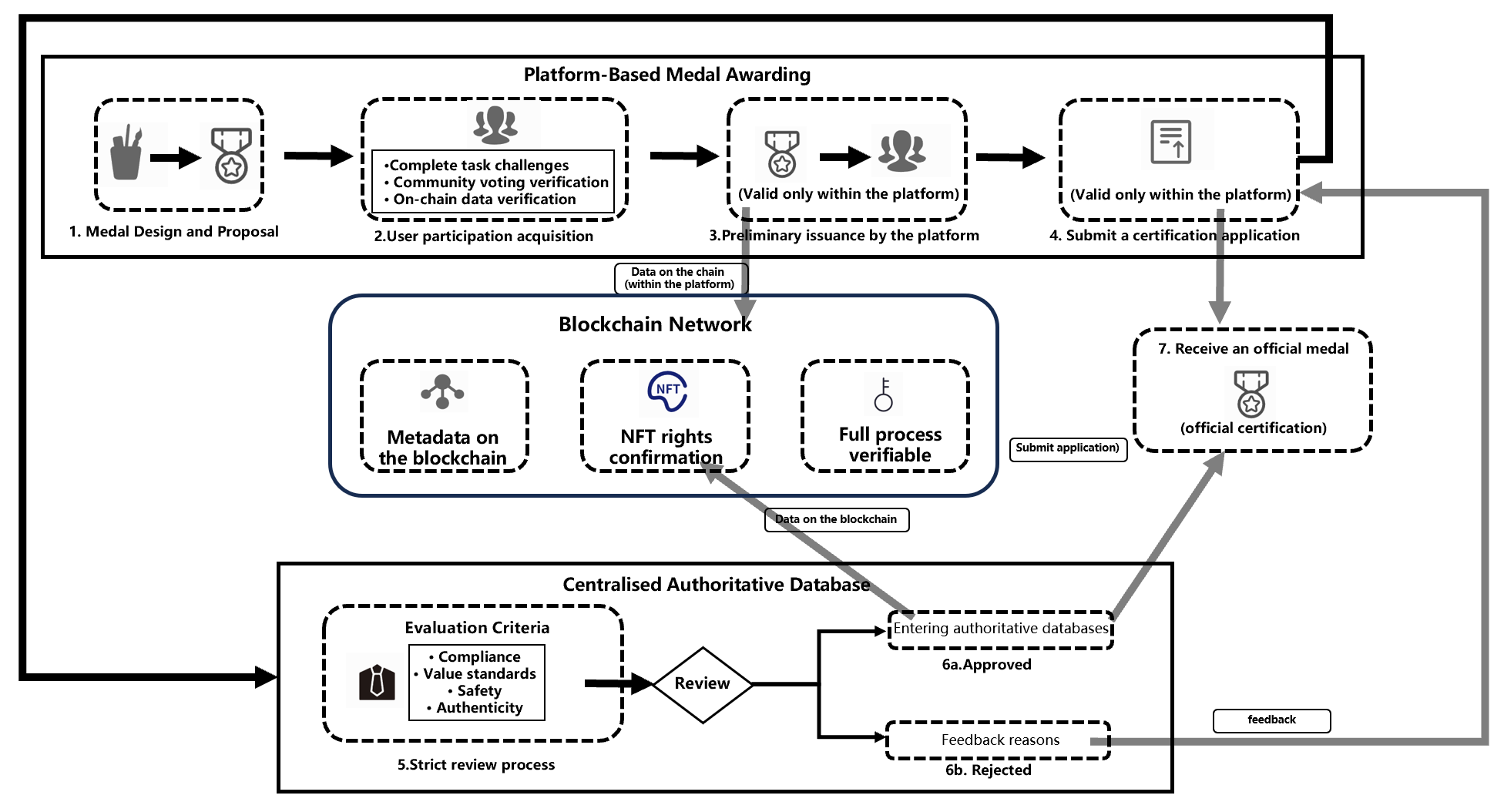}
    \caption{Detailed flowchart of medal certification process}
    \label{fig:placeholder}
\end{figure}

\subsection{System Advantages}

A centralised authoritative database ensures that the state maintains control over the content, standards, and final certification authority of medals, ensuring compliance with policies governing medal certification and issuance. This significantly enhances the credibility and social recognition of medal certification, strengthening its authority and trustworthiness. Decentralised platform operations alleviate the heavy workload of medal design and issuance, enabling the system to efficiently cover a wide range of fields and user groups, thereby improving the efficiency and scale of medal issuance.

The centralised review mechanism serves as the final safety valve to ensure the healthy operation of the medal system. which strictly screens out non-compliant content to ensure the high-quality operation of the entire ecosystem and maintain the system's high standards. The decentralised platform operation model encourages widespread user participation, allowing multiple platforms to jointly participate in the design, issuance, and operation of medals. This model not only stimulates the diversity and innovation of the ecosystem but also meets the diverse needs of different user groups, thereby unleashing more vitality and creativity. Mechanisms such as community voting and task challenges further enhance user engagement and a sense of belonging to the community. The application of blockchain technology ensures the secure storage of certification badge data. Through unique identifiers, it achieves data authenticity, immutability, and proof of ownership, providing robust technical support. 
 \begin{figure}[htbp]
     \centering
     \includegraphics[width=1\linewidth]{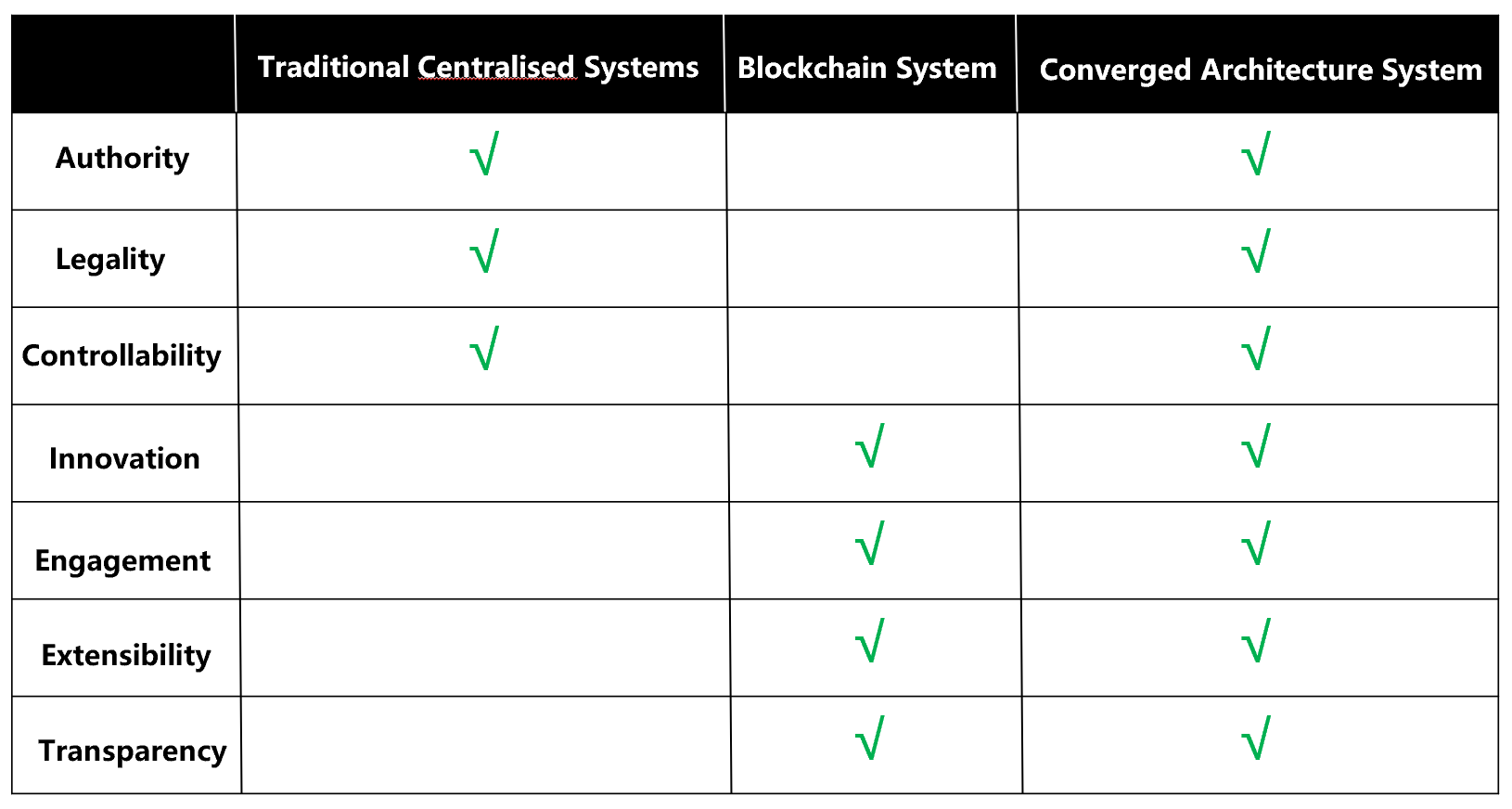}
     \caption{Comparison of the advantages of integrated architecture systems }
     \label{fig:placeholder}
 \end{figure}

\section{CONCLUTION}

Blockchain technology, as a key topic in the digital age, is being integrated into various projects with great enthusiasm. How to fully leverage blockchain-related technologies while meeting national policy requirements is crucial to the success of our system. The architecture of the blockchain-based electronic badge system cleverly balances the contradiction between {centralised control} and {decentralised vitality}. The centralised authoritative database serves as the foundation of trust and the final gatekeeper for certification, ensuring compliance, security, and the highest authority of the badges.The two are connected through a rigorous review and certification process: the platform creates the medals and awards them to users who meet the medal acquisition standards, which are then officially certified through strict review by the central authority and recorded in the authoritative database on the blockchain. This integrated model not only meets national policy requirements but also fully leverages the advantages of blockchain technology and community participation, thereby constructing a secure, trustworthy, vibrant, and sustainable electronic medal ecosystem. 
\section*{ACKNOWLEDGMENT}
The author would like to express sincere gratitude to Prof. Yi Xu for his invaluable guidance and support during the foundational stages of this research.

\bibliographystyle{ieeetr}
\bibliography{sample}
\end{document}